\documentclass[12pt]{article}

\usepackage{graphicx}
\usepackage{cite}
\usepackage{epsf}

\newcommand{\p}{{\bf p}}
\newcommand{\q}{{\bf q}}
\renewcommand{\r}{{\bf r}}
\renewcommand{\k}{{\bf k}}

\begin{document}

\begin{titlepage}
\begin{flushright}
{September 1999}
\end{flushright}
\vskip 0.5 cm

\begin{center}
            {\large{\bf Finite Temperature Time-Dependent 
 Effective Theory For The Goldstone Field In A BCS-Type Superfluid }}

\vspace{0.6cm}

 Ian J.R. Aitchison \footnote{E-mail address:
i.aitchison1@physics.oxford.ac.uk}, Georgios Metikas
\footnote{E-mail address: g.metikas1@physics.oxford.ac.uk} \\

\vspace{0.3cm}

Department of Physics, Theoretical Physics,\\
University of Oxford, 1 Keble Road, Oxford OX1 3NP

\vspace{0.3cm}

and \\
Dominic J. Lee \footnote{E-mail address: phydjl@phys-irc.novell.leeds.ac.uk}

\vspace{0.3cm}

Department of Physics,\\
University of Leeds, Leeds LS2 9JT
        
\end{center}

\vspace{2cm}

\abstract{We extend to finite temperature the time-dependent 
effective theory for the Goldstone field (the phase of the pair field)
 $ \theta $ which is appropriate for a superfluid containing one species of
 fermions with s-wave interactions, described by the BCS Lagrangian. We show that, when Landau damping is neglected, the effective theory can 
be written as a local time-dependent non-linear Schr\"{o}dinger Lagrangian 
(TDNLSL) which preserves the Galilean invariance of the zero
temperature effective theory and is identified with the superfluid
component. We then calculate the relevant Landau terms which are
non-local and which destroy the Galilean invariance. 
We show that the retarded $\theta$-propagator (in momentum space) can be well 
represented by two poles in the lower-half frequency plane, describing 
damping with a predicted temperature, frequency and momentum dependence. It is argued that the real parts of the Landau terms can be approximately interpreted as contributing to the normal fluid component.}

\end{titlepage}

\setcounter{footnote}{0}
 
\section{Introduction}

 The form of the effective (Ginzburg-Landau) Lagrangian for
time-dependent superconducting phenomena continues to attract attention. At $T=0$ Schakel \cite{Schakel}, using derivative expansion techniques \cite{Fraser}, obtained a time-dependent effective Lagrangian for the Goldstone field ( which is the phase $\theta$ of the pair field $\Delta$ ), starting from a simple BCS Lagrangian. Interestingly, Schakel's effective Lagrangian is the same as one that had been proposed earlier \cite{Witten} by assuming that the momentum density in a Galilean invariant system is proportional to a conserved current. 
In \cite{Thouless} (see also \cite{Stone}) the Lagrangian of
\cite{Schakel} and \cite{Witten} was shown to be equivalent, up to a
certain order in derivatives, to a time-dependent non-linear
Schr\"{o}dinger Lagrangian (TDNLSL); in particular, the role of
Galilean invariance was emphasized. More recently, de Palo et at. \cite{DiCastro} have given
a more general derivation of the TDNLSL at $T=0$, by including in the
effective action from the outset fluctuations in the density field
$\rho $ (conjugate to $\theta $), as well as those in $\Delta $. This
formulation allows a unified description of the weak-coupling (BCS)
and the strong-coupling (Bose) limits as well as of the cross-over
between them \cite{DiCastro}.   

In all the above cases, at $T=0$, electromagnetic interactions can be
straightforwardly included by the ``minimal coupling'' prescription,
since the effective Lagrangian is local. As is well-known \cite{Schrieffer},
 the $q^2=0$ pole of the $\theta$-mode is then pushed to finite
frequencies, becoming the plasma mode. 

At finite temperature, in the non-static case, the situation is
complicated by the presence of Landau damping terms, which appear to
prevent {\em any} local time-dependent Lagrangian for $\theta$, since
they have singularities at $q_{0}= \pm v_{F} |\q | $ (see Section 6
below), and the resulting cut structure implies that there is no
Taylor series expansion about the origin in momentum space. This
problem was first pointed out by Abrahams and Tsuneto
\cite{Tsuneto}. It poses a fundamental difficulty for the inclusion of
electromagnetism, since minimal coupling cannot be used unless the
effective Lagrangian is local. Note that the singularities, and hence
this difficulty, disappear in the static case, so that there is, of
course, nothing wrong with the familiar static Ginzburg-Landau theory
at $T \neq 0$.

Adopting a modern field-theoretical approach to the problem of
deriving an effective Lagrangian, Stoof \cite{Stoof} neglected the
Landau damping terms entirely and constructed a time-dependent
effective theory by writing the pair field as $\Delta ({\bf x},t)=
|\Delta_{0}| + \Delta^{\prime}({\bf x},t)$, where $|\Delta_{0}|$ is
the value of $\Delta$ which minimizes the effective action, and
$\Delta^{\prime}/|\Delta_{0}|$ was assumed to be varying slowly in
space and time. In \cite{Dominic} an attempt was made to extend
Stoof's approach so as to include Landau damping, retaining only terms
quadratic in $\Delta^{\prime}$. It was argued that the troublesome
(Landau) cut structure could be replaced - in the $\theta$-propagator
- by two simple poles at complex energy. This led to an approximate
 {\em local} effective action for $\theta$ of the form  

\begin{equation}
 \int d^4 x \left[ \dot{ \theta } ^2 - [ a(T) - i b(T) ] v_F^2 (
\nabla \theta )^2 \right] 
\label{leffaction} 
\end{equation}

\noindent where $ a(T)$, $b(T) $ are real functions. If
(\ref{leffaction}) were correct, minimal coupling could, as usual, be
employed to include electromagnetism. 

Unfortunately (\ref{leffaction}) is {\em not} correct. The imaginary
part of the Goldstone mode propagator, $S_{G}$, is an odd function of
the frequency and will not contribute to the total effective action as
in (\ref{leffaction}) (it does, however, enter the equation of motion
for $\theta$). The correct approximate two-pole form of $S_{G}$ will
be given in Section 7 below, and it will be such that $S_{G}^{-1}$ is
not local in coordinate space. Thus we are back with the problem of
how to include electromagnetism in the non-static case for $T \neq 0$.

Since we do not at present know of any answer to this problem, all our
subsequent analysis will generally refer to BCS-type pairing of {\em neutral}
fermions, i.e. to an uncharged fermionic superfluid. Helium-3 is
such a fluid but it condenses into a p-wave state, not the s-wave BCS
state studied here. It would be straightforward, and interesting, to
generalize the present approach to p-wave pairing (a beginning was
made by Van Weert \cite{VanWeert}). However, it is likely that trapped
condensates of neutral fermionic fluids will be produced before very
long (see for example \cite{Houbiers}). Our analysis would be relevant to such a case, when re-done for a
finite system.

Apart from the error in $S_{G}$, the approach leading to
(\ref{leffaction}) is unsatisfactory for another reason, namely it
fails to reduce, at $T=0$, to the correct Galilean-invariant result of
\cite{Schakel,Witten,Thouless}. The problem can be traced to the step
$\Delta=|\Delta_{0}|+\Delta^{\prime}$: as was pointed out in
\cite{Thouless}, this ansatz destroys Galilean invariance and fails to
allow for large fluctuations in $\theta$, two weaknesses which are in
fact related. In \cite{Dominic}, the further approximations $\Delta
\approx |\Delta| \ (1+i\theta)$ and $ |\Delta|=\mathrm{constant}$ were
made, which led to the form (\ref{leffaction}) after retaining only
quadratic terms in $\mathrm{Im}(\Delta)$. What is required, rather, is
the finite temperature version of \cite{Thouless} (or even better of
\cite{DiCastro}), where we write $\Delta=|\Delta| \ e^{i \theta}$,
ignore fluctuations in $|\Delta|$ and consider a systematic expansion
in the derivatives of $\theta$, without having to assume that $\theta$
itself is small. 

Our aim in this paper is to provide the finite temperature extension
of \cite{Thouless} and arrive at a temperature and time-dependent
effective Lagrangian for $\theta$, $L(\theta)$. In Section 2 we set up the
effective action formalism and derive the usual gap equation.
 In Section 3 we evaluate the first term in the expansion of the
effective action in powers of derivatives and in Section 4 we
calculate the next term which includes the Landau damping
contribution. In Section 5 we show that, if the Landau terms are
neglected, Galilean invariance is maintained (at least up to the
order in derivatives to which we are working) and $ L(\theta ) $ has
exactly the same form as in \cite{Schakel, Witten, Thouless}
 but with temperature dependent coefficients. This, in turn, implies that an
effective non-linear Schr\"{o}dinger theory can be obtained with a
temperature-dependent potential term. 

Next, we turn our attention to the Landau terms, which are calculated
 in momentum space in Section 6. As expected, their form is similar to the
expressions found in \cite{Dominic}. Their presence prevents any representation
of the effective action in terms of a local Lagrangian. In Section 7 we consider how to obtain a useful approximate
 form for the equation of motion for $\theta$, including the Landau
terms. Noting (see above) that $\mathrm{Im}(S_{G})$ is an odd function
of frequency, we show that $S_G$ can be well represented by 
two poles in the lower half plane of the complex frequency
variable. This corrects the form proposed 
in \cite{Dominic} and invalidates (\ref{leffaction}).  
We then consider the real parts of the Landau terms, neglecting the
imaginary parts. We propose that they can be approximated by their value at the (undamped) poles of 
$S_G$ - an ``on-shell'' approximation. The complete effective action then 
becomes local, though the Landau terms break Galilean invariance. We propose 
a physical interpretation of the resulting equation of motion, in which the 
real parts of the Landau terms represent a contribution to the normal fluid 
component.

\newpage

\section{The Effective Action and the Gap Equation}

We begin with a brief review of the formalism. The BCS Lagrangian, for s-wave pairing and in the absence of external fields, is
 
\begin{equation}
 L = \sum_{\sigma=1}^{2} \psi^*_{\sigma} (x) ( i  \partial_t + \frac{\nabla^2}{2m} + \mu ) \psi_{\sigma} (x) + g \psi^*_{\uparrow} (x) \psi^*_{\downarrow} (x) \psi_{\downarrow} (x) \psi_{\uparrow} (x) 
\end{equation} 

\noindent where $\psi_{\sigma} (x)$  describes electrons with spin
$\sigma = ( \uparrow , \downarrow )$, $ \mu = \frac{k_F^2}{2m}$  is
the Fermi energy, $x=(t, {\bf x})$ and $g>0$.  We define the Nambu spinor 

\[ \psi = \left( \begin{array}{c}               
          \psi_{\downarrow} \\
          \psi^*_{\uparrow} 
          \end{array} \right) \]

\noindent and introduce the auxiliary ``pair'' fields $\Delta(x)$ and 
$\Delta^{*}(x)$ so as to write the partition function as 

\begin{equation}
 Z = \int D \psi^{\dagger} D \psi \int D \Delta^* D \Delta \ \exp
\left[ i \int
d^4x \ ( \psi^{\dagger} { \cal G }_{xx}^{-1} \psi
- \frac{ | \Delta |^2}{g} ) \right]
\end{equation}

\noindent where \[ { \cal G }_{xx}^{-1} = \left( \begin{array}{cc}
                  i \partial_t + \frac{\nabla^2}{2m} + \mu & \Delta \\
                  \Delta^* &  i \partial_t - \frac{\nabla^2}{2m} - \mu
                  \end{array} \right). \] 

\noindent We now do a gauge transformation on the electron field
 \cite{Schakel}, writing
$ \psi_{\sigma} = \exp \left( \frac{i \theta (x) }{2} \right)
\chi_{\sigma} (x) $, where $\theta$ is the phase of $\Delta$ and $ \chi_{\sigma}$ are new anticommuting fields.
The integrals over $\chi$ and $\chi^{\dagger}$ are then performed,
leading to 
 
\begin{equation}
Z = \int D \Delta^* D \Delta \ \exp{ (i S_{\mathit{eff}}[\Delta,\Delta^{\dagger}]) } 
\end{equation}

\noindent where the effective action $S_{\mathit{eff}}$ is given by 
 
        \begin{eqnarray}
                  S_{\mathit{eff}}[\Delta,\Delta^*] & = & -i \mathrm{Tr} \log{ ( \hat{G}^{-1} -\hat{\Sigma } ) } - \int d^4x \frac{ |\Delta |^2}{g} \\
           & = & -i \mathrm{Tr} \log{ \hat{G}^{-1}} + i \mathrm{Tr} \sum_{n=1}^{\infty} \frac{(\hat{G}\hat{\Sigma})^n}{n}-
 \int d^4x \frac{ |\Delta|^2 }{g}  \label{seffseries} 
        \end{eqnarray}

\noindent with 

         \begin{eqnarray*}
                                \hat{G}^{-1} & = & \left( \begin{array}{cc}
                  \hat{p}_o - \frac{ \hat{ \p\ }^2}{2m} + 
\mu & | \hat{\Delta} | \\
                | \hat{\Delta} | & \hat{p}_0 + \frac{\hat{ \p\ }^2}{2m} - \mu 
                  \end{array} \right)  \\
                 \hat{ \Sigma } & = & \left( - \frac{i \widehat{ \nabla^2
                                 \theta}}{4m} + \frac{ ( \widehat{\nabla\theta})
                                 \hat{\p\ } }{2m} \right) I + \left(
                                 \frac{ \widehat{ \dot{\theta} } }{2} + \frac{
                                  \widehat{ \left( \nabla \theta \right)^2}}{8m}
                                 \right) \tau_3   \\
                  \hat{f} & = & f ( \hat{x} ). 
\end{eqnarray*}

\noindent Note that the effect of the gauge transformation has been to 
separate the dependence on $|\hat{\Delta}|$ which is in $\hat{G}^{-1}$
from that on $\hat{\theta}$ which is in $\hat{\Sigma}$. We point out that $ \mathrm{Det} $ and $ \mathrm{Tr} $ refer to both the
functional space and the internal (Nambu) space. The operators
 $ \hat{p_0} $, $ \hat{\p\ } $ and $ \hat{x} $ are Fock-Schwinger
Proper Time operators. 

The ground state is determined by taking $|\Delta|$ to be space-time 
independent and solving the mean field equation

\begin{equation}
\frac{\delta S_{\mathit{eff}}}{\delta | \Delta |} = 0 \Leftrightarrow
ig \int \frac{d^4p}{(2 \pi)^4} \frac{1}{p_0^2 - E^2} = 1   \label{mean
field eq}
\end{equation}

\noindent where $ E^2 = \epsilon^2 + | \Delta |^2 $ and $ \epsilon =
\frac{ \p\ ^2 }{2m} - \mu $ . The modulus $| \Delta |$ of the pair field
is therefore the {\em energy 
gap}. (\ref{mean field eq}) can be written in the usual form of the
BCS gap equation at finite temperature \cite{Bardeen, Schrieffer} 

\begin{equation}
g \int \frac{d^3 \p\ }{ (2 \pi)^3 } \frac{1}{2E} \
\tanh{  \frac{\beta E}{2} } = 1
\end{equation}

\noindent where we assume that the potential (the coupling constant)
is attractive in a shell of width $ 2 \omega_c $ centered at the Fermi
surface and vanishes outside this shell. 
  We note that in the weak-coupling limit $ \omega_c \gg | \Delta(0) |  $, where $| \Delta(0)|$ refers to the value of the gap within the shell.
 In this limit we can consider $ | \Delta |$ to be a
 time and space independent constant whose value is
   
\begin{equation}
|\Delta(T)| \simeq |\Delta(0)| \simeq 1.75 k_B T_c  \label{Delta}
\end{equation}

\noindent for 

\begin{equation}
b=T/T_c \in [0.1, 0.6].  \label{b}
\end{equation}

Thus the effective theory only involves the single degree of freedom $\theta$.
 If we expand each term of the infinite series in (\ref{seffseries})
in powers of momentum and frequency and restrict ourselves to small
values of these quantities, it has been proved in \cite{Schakel, Thouless}
 that at $T=0$ the trace of the
first term contributes a term proportional to $ [ \dot{\theta} +
(\nabla \theta)^2 / 4m ] $ to $ L(\theta) $, and the trace
of the second term contributes a term proportional to $ [ \dot{\theta} +  (\nabla \theta)^2 /4m]^2 $. As emphasised in \cite{Thouless}, the combination 
$ [ \dot{\theta} + (\nabla \theta)^2 /4m] $ is Galilean invariant and it is 
this feature which allows $L(\theta)$ to be rewritten as a TDNLSL
 at zero temperature. We shall now  
extend this calculation to finite temperature and explore to what extent 
an effective Lagrangian is still derivable and - if so - whether it can be 
written as a TDNLSL.

\section{The First Trace}

In this section we calculate the first $(n=1)$ term of the sum appearing in (\ref{seffseries}) :

\begin{equation}
\mathrm{Tr} [ \hat{G} \hat{\Sigma} ] = 
\int  \frac{d^4p}{(2 \pi)^4} \ \mathrm{tr} [ G(p_0,\p\ ) \tau_3 ] \int
d^4x \  \frac{1}{2} \ [ \dot{\theta} + \frac{ (\nabla \theta)^2 }{4m}]\label{first trace}
\end{equation}

\noindent where a term odd in $p_0$ has vanished in the last step,
since we integrate over $p_0$ from $- \infty$ to $+ \infty$.
By $ \mathrm{tr} $ we mean the trace only in the internal (Nambu) space.
We apply the Matsubara formalism to
evaluate the first integral at finite temperature. 
 We Wick-rotate  via $p_{0} \rightarrow i p_{0E}$ and then replace 
the integral over $p_{0E}$ by a sum over discrete fermion 
frequencies via  
\begin{equation}
i \int_{-\infty}^{\infty} \frac{dp_{0E}}{2 \pi} f(i p_{0E}) \rightarrow 
\frac{i}{\beta} \ \sum_{n=-\infty}^{\infty} f( i z_n) \label{discretisation}
\end{equation}

\noindent where $z_{n}=(2n+1) \pi / \beta $, and 
 $n$ is integer. Next, following \cite{Zuk}, we write 

\begin{equation}
\frac{i}{\beta} \sum_{n} f(i z_n) \equiv \frac{i}{\beta}
\sum_{n} f_E (z_n) = \frac{-1}{4 \pi
} \ \int_{C_R} dz \ f_E(z) \ \tan{ ( \frac{ \beta z }{2} )}  \label{fTsum}
\end{equation}

\noindent where the contour $C_R$ encloses the real axis in a counter-clockwise
 sense. (\ref{fTsum}) is valid provided that $f_E(z)$ has no
singularities on the real axis, which is our case. It is convenient to
rewrite the above equation in such a way as to separate out the (known) $T=0$ contribution :

\begin{equation}
\frac{-1}{4 \pi} \ \int_{C_R} dz \ f_E(z) \ \tan{ ( \frac{ \beta z
}{2} ) } = L1 + L2 + L3 \label{fTsum1} 
\end{equation}

\noindent with \\

\begin{eqnarray}
L1&=& \frac{i}{2 \pi} \int_{ - \infty }^{ \infty } dz \ f_E(z)
\nonumber \\
L2&=& - \frac{i}{2 \pi} \int_{ - \infty + i \eta}^{ \infty + i
\eta} dz \ f_E(z) \  n(-iz) \nonumber \\
L3&=& - \frac{i}{2 \pi} \int_{ - \infty -i
\eta}^{ \infty -i \eta} dz \ f_E(z) \ n(iz) \nonumber \\ 
\end{eqnarray}

\noindent where $ \eta $ is an infinitesimal positive
quantity and $ n(x)= \frac{1}{1 + \exp{
\beta x}} $ is the Fermi-Dirac distribution which has been introduced
in our calculation through the use of the identities  $ \tan{( \frac{\beta z }{2})}=-i[1-2n(iz)]=i[1-2n(-iz)] $.

Applying the above formalism, we find that the total contribution of
the first term of the series for $ S_{\mathit{eff}} $ is
 
\begin{equation} 
i \mathrm{Tr} [\hat{G} \hat{\Sigma}] = - \int  \frac{d^3 \p\ }{(2 \pi)^3}
 \left[ 1 - \frac{\epsilon}{E} [1-2 n(E)] \right]
 \int d^4x \  \frac{1}{2} \ [ \dot{\theta} + \frac{ (\nabla \theta)^2
}{4m}] 
\label{1trace}
\end{equation}

\noindent which has the same Galilean-invariant form as at $T=0$ but
with a temperature dependent coefficient. In fact, the temperature dependence of the first
trace is not important and can be neglected. The maximum value
of the function $n(E)$ for the range of
temperatures of (\ref{b}) and with the gap having the value given in
 (\ref{Delta}) is $ n_{max} \simeq 0.05 $.  
Consequently $ 1-2n(E) \simeq 1 $ and the coefficient of the Galilean invariant term  $ [ \dot{\theta} +
\frac{ (\nabla \theta)^2 }{4m} ] $ is 
\begin{equation}
 M(T) = \frac{1}{2} \ \int  \frac{d^3\p\ }{(2 \pi)^3}
 \left[ 1 - \frac{\epsilon}{E} [1-2 n(E)] \right] \simeq \frac{1}{2} \
 \rho_0 = M(0),   \label{M}   
\end{equation}
where $\rho_{0}=p_{F}^{3}/(3 \pi^2)$ is the electron density at $T=0$.

\section{The Second Trace}

The trace of the second term $(n=2)$ of the infinite series in (\ref{seffseries}) can be
written as:
\begin{equation}
\hspace*{-2cm} \mathrm{Tr} [\hat{\Sigma} \hat{G} \hat{\Sigma} \hat{G}]
 = \mathrm{tr} \int \frac{d^4p}{(2\pi)^4} \frac{d^4q}{(2\pi)^4} \ \tilde{\Sigma}(q,\p\ -
\frac{\q\ }{2} ) G(p_0-\frac{q_0}{2},\p\  - \frac{\q\ }{2} ) \tilde{\Sigma}(-q,\p\
+\frac{\q\ }{2}) G(p_0 + \frac{q_0}{2}, \p\ +\frac{\q\ }{2}) 
\end{equation}

\noindent where $\tilde{\Sigma} (p,\q\ )=\int d^4x \ \exp{(ip.x)} \
\Sigma(x,\q\ )$. After performing the Wick rotation we have 
\begin{equation} 
\hspace*{-2cm} \mathrm{Tr} [ \hat{\Sigma} \hat{G} \hat{\Sigma} \hat{G}]_E = - \int \frac{d^4q_E}{(2\pi)^4} \frac{d^4p_E}{(2\pi)^4} T_1 - 2i \int \frac{d^4q_E}{(2\pi)^4}
\frac{d^4k_E}{(2\pi)^4} \frac{d^4p_E}{(2\pi)^4} T_2 + \int \frac{d^4q_E}{(2\pi)^4} \frac{d^4k_E}{(2\pi)^4} \frac{d^4r_E}{(2\pi)^4} \frac{d^4p_E}{(2\pi)^4} T_3  \nonumber 
\end{equation}

\noindent where
 
\begin{eqnarray*}
 &&\hspace*{-2cm} T_1 = \tilde{\theta}_E (q_E) \
\tilde{\theta}_E (-q_E) \ A_{E-} \ A_{E+} \left[ B_E [\frac{( \p\ .\q\
)^2}{2m^2} + \frac{q_{0E}^2}{2}] + C_E [- \frac{( \p\ .\q\
)^2}{2m^2} +\frac{q_{0E}^2}{2}]
+ D_E q_{0E} \frac{ \p\ .\q\ }{m} \right]  \\
 &&\hspace*{-2cm} T_2 = \tilde{\theta}_E(q_E) \
\tilde{\theta}_E(k_E) \ \tilde{\theta}_E(-q_E-k_E) \ A_{E-} \
A_{E+}  \frac{-(\k\ .\q\ + \k\ ^2) }{8m} \left[[B_E+C_E]\
 q_{0E} + D_E \frac{ \p\ .\q\ }{m} \right]  \\ 
 &&\hspace*{-2cm} T_3 = \tilde{\theta}_E(k_E) 
 \tilde{\theta}_E(r_E) \ \tilde{\theta}_E(q_E-k_E) \
\tilde{\theta}_E(-q_E-r_E) \ A_{E_-} \ A_{E_+} \frac{-\k\ ^2+ \k\ .\q\ }{32 m^2} ( \r\ ^2 + \r\ .\q\ ) [ B_E + C_E ] \nonumber   
\end{eqnarray*}
\noindent with 
\begin{eqnarray*}
&&\hspace*{-2cm} A_E(p_E)= \frac{-1}{p_{0E}^2+E(\p\ )^2},\ A_{E+}=A_E(p_E+\frac{q_E}{2}),\ A_{E-}=A_E(p_E-\frac{q_E}{2})\\
&&\hspace*{-2cm} B_E(p_{0E},q_{0E},\p\ ,\q\ )= |\Delta|^2 \\
&&\hspace*{-2cm} C_E(p_{0E},q_{0E},\p\ ,\q\ ) = - \epsilon(\p\ -\frac{\q\
}{2}) \epsilon(\p\ +\frac{\q\ }{2}) + (p_{0E}^2-(\frac{q_{0E}}{2})^2) \\
&&\hspace*{-2cm} D_E(p_{0E},q_{0E},\p\ ,\q\ ) = \epsilon(\p\ + \frac{\q\ }{2}) (p_{0E}- \frac{q_{0E}}{2}) + \epsilon(\p\ -\frac{\q\ }{2}) (p_{0E} + \frac{q_{0E}}{2}).
\end{eqnarray*}

\noindent We note that the integrals over $p_{0E}$ all have the form

\begin{equation}
\tilde{g}_E=\int_{-\infty}^{\infty} \frac{dp_{0E}}{2\pi}\  A_{E-}\ A_{E+}\ g_E
(p_{0E},q_{0E},\p\ ,\q\ ) \hspace{1.2cm} g=B, C, D .   \label{gtilde}
\end{equation}

\noindent We now proceed to the second step of the finite temperature
 formalism which is to discretise the momentum variables as follows

\begin{eqnarray}
p_{0E} \rightarrow p_{0E}^n=\frac{(2n+1)\pi}{\beta}&,& q_{0E}/2
\rightarrow q_{0E}^m/2=\frac{(2m\pi)}{\beta} \nonumber \\ 
k_{0E} \rightarrow k_{0E}^l=\frac{(2l\pi)}{\beta}&,& r_{0E}
\rightarrow r_{0E}^s=\frac{(2s\pi)}{\beta} \label{discretisation2}
\end{eqnarray}

\noindent and then calculate the sums over $n$ which are contained in
 (\ref{gtilde}) by replacing them with integrals over $z$ according to (\ref{fTsum}).
The poles of $ \tan(\frac{\beta z}{2}) $ are on the real axis and the
sum of their residues is equal to the finite temperature sum which is 
to be calculated. We note that only $A, B, C, D$ depend on the variable
$z$ and $B, C, D$ have no poles. This means that we need to evaluate only the following integrals
\begin{equation}
\hspace*{-1cm} \tilde{g}_E = \int_{C} dz \ A_E(z + \frac{q_{0E}^m}{2}, \p\ + \frac{\q\ }{2} )\
A_E(z - \frac{q_{0E}^m}{2},\p\ - \frac{\q\ }{2}) \ g_E(z,q_{0E}^m,\p\
,\q\ ) \ \tan{(\frac{\beta z}{2})} \label{integral}
\end{equation}

\noindent where the function $g_E$ has no poles (analytical everywhere in the
 $z$-plane) and the $A$-functions have the poles
\[ \frac{q_{0E}^m}{2} + i E_- ,\ \frac{q_{0E}^m}{2} - i E_- ,\
-\frac{q_{0E}^m}{2} + i E_+,\ -\frac{q_{0E}^m}{2} - i E_+ \]  where we
have introduced the usual notation 
\[  E_+ = E(\p\ + \frac{\q\ }{2}) \hspace{3cm} E_- =
E(\p\ - \frac{\q\ }{2}). \] 

We follow the same procedure as for the first trace.  
We note from (\ref{discretisation2}) that $ \exp{ (i \beta
\frac{q_{0E}^m}{2})} = 1$ and then, following \cite{Stoof, Schrieffer,
Zuk}, we Wick rotate the above expression back to real continuous
variables $q_0$ via $q_{0E}^m \rightarrow -i q_0$. The result can be written as
  $ \tilde{g} = \tilde{g}_R +
\tilde{g}_L$, where  the regular part  $\tilde{g}_R $
has a well-defined expansion about the origin $(q_0,\q\
)=(0,{\bf 0})$, but the Landau part  $\tilde{g}_L$
 does not, as was first pointed out in \cite{Tsuneto}. For $ g= B, C, D $ we have 

\begin{eqnarray*}
\tilde{B}_R=&& \frac{|\Delta|^2}{4 E_+ E_-}\
[1-n(E_+)-n(E_-)]\left( \frac{1}{E_+ + E_- -q_0} + \frac{1}{E_+ + E_-
+q_0} \right)  \nonumber \\
\tilde{B}_L=&&  \frac{|\Delta|^2}{2 E_+ E_-}  \frac{n(E_+) - n(E_-)}{E_+ - E_- +q_0}  \nonumber \\
\tilde{C}_R=&&  F_1 [1-n(E_+)-n(E_-)] \left( \frac{1}{E_+ + E_-  -q_0} +
\frac{1}{E_+ + E_- + q_0}\right) \nonumber \\
\tilde{C}_L=&& 2F_2 \frac{n(E_-) -n(E_+)}{E_+ - E_- + q_0} \nonumber \\
\tilde{D}_R=&& G_1\ [1-n(E_+)-n(E_-)] \left( \frac{1}{E_+ + E_- - q_0} -
\frac{1}{E_+ + E_- + q_0} \right) \nonumber \\
\tilde{D}_L= && 2G_2 \frac{n(E_+) - n(E_-)}{E_+ - E_- +q_0}
\end{eqnarray*}

\noindent with
\[ F_1=\frac{1}{4} \left[1-\frac{\epsilon_+ \epsilon_-}{E_+ E_-}
\right],\  
   F_2=\frac{1}{4} \left[1+\frac{\epsilon_+ \epsilon_-}{E_+ E_-}
   \right] \] \\
\[ G_1=\frac{i}{2} \frac{\epsilon_+}{E_+},\ 
G_2=\frac{i}{4} \left[ \frac{\epsilon_+}{E_+} + \frac{\epsilon_-}{E_-} \right]. \]

\noindent After the analytic continuation, wherever we write $q_0$ in
the denominators of $B,C,D$, we imply that there is an
infinitesimal imaginary part, $ i \, \delta $, $\delta \rightarrow
0^+$ added to $ q_0 $, corresponding to retarded boundary conditions
\cite{Schrieffer,WeldonMishaps,Stoof}. We complete the Wick rotation
back to continuous variables via
$k_{0E}^m \rightarrow -ik_0,\ r_{0E}^m \rightarrow -ir_0$, and the
second trace finally takes the form

\begin{eqnarray}
 \mathrm{Tr} [ \hat{\Sigma} \hat{G} \hat{\Sigma} \hat{G}]  &=
 & i \int \frac{d^4q}{(2\pi)^4} \tilde{\theta} (q) \
\tilde{\theta} (-q) \int \frac{d^3\p\ }{(2\pi)^3} \left[ \tilde{B} [ \frac{(\p\
.\q\ )^2}{2m^2} - \frac{q_0^2}{2} ] \right. \nonumber \\
 & & \left. + \tilde{C} [ - \frac{(\p\ .\q\
)^2}{2m^2}  - \frac{q_0^2}{2} ] + \tilde{D} \ (-i)q_0 \frac{\p\ .\q\ }{m} \right] \nonumber \\
 &\hspace{-1cm} + 2i & \hspace{-0.5cm}  \int \frac{d^4q}{(2\pi)^4}
\frac{d^4k}{(2\pi)^4} \tilde{\theta} (q)\ \tilde{\theta} (k) \tilde{\theta} (-q
- k) \frac{-(\k\ .\q\ +\k\ ^2)}{8m} \nonumber \\
& & \int \frac{d^3\p\ }{(2\pi)^3} \left[ [ \tilde{B} +
 \tilde{C} ] (-i) q_0 + \tilde{D} \frac{\p\ .\q\ }{m} \right] \nonumber \\
& \hspace{-1cm} +i & \hspace{-0.5cm} \int \frac{d^4q}{(2\pi)^4} \frac{d^4k}{(2\pi)^4} \frac{d^4r}{(2\pi)^4}
\tilde{\theta} (k)\  \tilde{\theta} (r) \ \tilde{\theta} (q - k) \
\tilde{\theta} (-q -r) \nonumber \\ 
& & \frac{(-\k\ ^2 +\k\ .\q\ )(\r\ ^2 + \r\ .\q\ )}{32m^2} \int
\frac{d^3\p\ }{(2\pi)^3} \ \left[ \tilde{B} + \tilde{C} \right]. \label{2trace}
\end{eqnarray}

\section{The effective Lagrangian at $T \not= 0$, neglecting the Landau terms}

The contribution of the first Trace to the effective action is given in (\ref{1trace}). 
 When the Landau terms are neglected, only the functions $\tilde{B}_R, \tilde{C}_R$ and $\tilde{D}_R$ appear in the expression (\ref{2trace}) for the second Trace and they can all be Taylor-expanded about $q=0$ in powers of $q_0/\mid \Delta \mid$
 and $v_F \mid {\bf q} \mid / \mid \Delta \mid$, so long as these quantities are small.  For $\tilde{B}_R$ and $\tilde{C}_R$ the expansion takes the form

\begin{eqnarray}
 \tilde{B}_R ({\bf p},q) &=& \frac{\mid \Delta \mid^2}{4E^3} (1-2n(E)) + O({\bf q}^2, ({\bf q} . {\bf p})^2, q^2_0)  \label{BR}\\
\tilde{C}_R ({\bf p }, q) &=& \frac{\mid \Delta \mid^2}{4E^3} (1-2n(E)) + O({\bf q}^2, ({\bf q}. {\bf p})^2, q^2_0).  \label{CR}
\end{eqnarray}

\noindent $\tilde{D}_R(q)$, on the other hand, vanishes as $q=0$, and we have

\begin{equation}
 \tilde{D}_R({\bf p}, q) = O(q_{0}({\bf q}.{\bf p})). \label{DR}\\
\end{equation}

\noindent When (\ref{BR}) - (\ref{DR}) are inserted in (\ref{2trace}) and the appropriate Fourier transforms performed, we obtain a local effective Lagrangian for $\theta$, involving powers and products of the derivatives $\theta$ and $\nabla \theta$.

We must now consider how this derivative expansion is to be ordered.
Due to the preferred frame supplied by the heat bath, we do not expect
Galilean invariance to be preserved in the effective Lagrangian at $T
\not= 0$.  Nevertheless, we expect Galilean invariance to remain
approximately valid for $T \leq T_c$.  Accordingly we shall regard
terms of the form ``$ \dot{\theta} $'' and ``$(\nabla \theta)^2/m$''
as being of the same order - or equivalently, terms of the form ``$q_0\theta$'' and ``${\bf q}^2
 \theta^2$'' as the same order.  It then follows that the leading order terms in the expansion of 
(\ref{2trace}) are given by retaining only the {\em zeroth} order contributions to $\tilde{B}_R, \tilde{C}_R$ and $\tilde{D}_R$ from 
(\ref{BR})-(\ref{DR}), thanks to the explicit factors involving $q_0$ and ${\bf q}$ already appearing in (\ref{2trace}). This is an important point.\\

\noindent In evaluating these leading order terms, it is convenient to define

\begin{equation}
 N(T) = \int \frac{d^3{\bf p}}{(2 \pi)^3} \ \frac{\mid \Delta \mid^2}{2E^3} (1-2n(E)),
\end{equation}

\noindent and to make use of the approximation \cite{Stoof}

\begin{equation}
 N(T) \approx \frac{1}{2} N(0) R(T) \label{fermi}
\end{equation}

\noindent where

\begin{equation}
 N(0) = \frac{m p_F}{2\pi^2} = \frac{3 \rho_{0}}{2mv_{F}^2} 
 \label{N(0)}
\end{equation}
 
\noindent is the density of states at the Fermi energy and

\begin{equation}
R(T) = \int dy \frac{1-2n(y)}{(1+y^2)^{\frac{3}{2}}}
\end{equation}

\noindent where $y=\epsilon/\mid \Delta \mid$.
The leading contribution to the first term in (\ref{2trace}) is then

\begin{equation}
-i \frac{1}{2} N(T) \int \frac{d^4q}{(2\pi)^4} \ \tilde{\theta}(q)
\tilde{\theta}(-q)q^2_0 = -\frac{i}{2} N(T) \int d^4x \ \dot{\theta}^2(x).\\
\end{equation}

\noindent Evaluating the other terms in the same way, and including the expression
 (\ref{1trace}) for the first Trace, we obtain the effective Lagrangian (neglecting Landau terms)

\begin{equation}
{\cal{L}}_R(\theta) = -M(T) \left[ \dot{\theta} + \frac{(\nabla
\theta)^2}{4m} \right] + \frac{N(T)}{4} \left[ \dot{\theta} + \frac{(\nabla
\theta)^2}{4m} \right]^2. 
\label{LR} 
\end{equation}

\noindent Remarkably, Galilean invariance is still preserved by 
(\ref{LR}). The only difference between (\ref{LR}) and ${\cal{L}}_{\mathit{eff}} (\theta)$ of 
\cite{Schakel, Thouless} is the appearance of the $T$-dependent coefficients $M(T)$ 
and $N(T)$.  In fact, as $T \to 0$ we note that $M(T) \to \rho_{0}/2$,
while $N(T) \to N(0)$, so the $T=0$ result of \cite{Schakel, Thouless} is easily regained.

The Galilean invariance of (\ref{LR}) guarantees that it can be rewritten as a time-dependent non-linear Schr\"{o}dinger Langrangian (TDNLSL), as explained in
 \cite{Thouless}.  The results of \cite{Thouless}
 can be immediately taken over to $T\not= 0$ (always with the neglect of the Landau terms) by making the replacements $\rho_{0}/2 \to M(T), N(0) \to N(T)$. 
 The result is merely that the ``potential'' term in the equivalent TDNLSL becomes

\begin{equation}
 V=\frac{\left[ \rho - 2M(T) \right]^2}{2N(T)}.
\end{equation}

\noindent Thus, always with neglect of Landau terms, we have shown
that the effective theory for our neutral fermionic fluid is a TDNLSL,
just as it is for a bosonic fluid. The inclusion of electromagnetism
by means of the ``minimal coupling'' prescription is straightforward.

We now turn to the Landau terms which, as in \cite{Dominic}, are
non-local and cannot be Taylor expanded about $q=0$.

\section{The Landau terms}

The Landau terms $\tilde{B}_L,\tilde{C}_L$ and $\tilde{D}_L $
 cannot be treated in the same way as the regular
terms, because of their non-analyticity around the origin. However,
since we are interested in the low frequency, long wavelength regime,
we can follow \cite{Zuk,Tsuneto} and write 
\begin{equation}
E_+ - E_- = \frac{\epsilon}{E}\ \frac{\p .\q }{m}. 
\end{equation}
\noindent We now expand the numerators of these quantities in
powers of momentum  ${\q}$ and keep the leading order terms which are of
first order in $\q$. We find

\begin{eqnarray}
\tilde{B}_L & \approx & \left[ \frac{\epsilon}{E}\ \frac{\p .\q }{m}+q_0+
i \delta 
\right]^{-1}\ \frac{|\Delta|^2}{2E^2}\ \frac{dn}{dE} \
\frac{\epsilon}{E}\ \frac{\p .\q }{m} \nonumber \\
\tilde{C}_L &\hspace{-0.5cm} \approx  & - \left[ \frac{\epsilon}{E}\ \frac{\p .\q }{m}+q_0+ i \delta 
\right]^{-1}\ \frac{2\epsilon^2 + |\Delta|^2}{2E^2} \ \frac{dn}{dE} \
\frac{\epsilon}{E}\ \frac{\p .\q }{m} \nonumber \\
\tilde{D}_L & \approx & \left[  \frac{\epsilon}{E}\ \frac{\p .\q }{m}+q_0 +
i \delta  \right]^{-1} \
\frac{dn}{dE} \ \frac{i \epsilon^2}{E^2} \ \frac{\p .\q }{m}. \label{apLandau}
\end{eqnarray}

\noindent If we look at (\ref{2trace}) and recall that $\tilde{g} =
\tilde{g}_R + \tilde{g}_L $ for $\tilde{g}=\tilde{B}, \tilde{C},
\tilde{D}$ we realize that the Landau terms appear in the
second Trace in the following integrals 

\begin{eqnarray}
I_1 &=& \int \frac{d^3\p\ }{(2\pi)^3} \left[ \tilde{B}_L [ \frac{(\p\ .
\q\ )^2}{2m^2} - \frac{q_0^2}{2} ] + \tilde{C}_L [ - \frac{(\p\ .\q\
)^2}{2m^2}  - \frac{q_0^2}{2} ] + \tilde{D}_L \ (-i)q_0 \frac{\p\ .\q\
}{m} \right] \nonumber \\
I_2 &=& \int \frac{d^3\p\ }{(2\pi)^3} \left[ [ \tilde{B}_L + \tilde{C}_L ] (-i) q_0 + \tilde{D}_L \frac{\p\ .\q\ }{m} \right] \nonumber \\
I_3 &=& \int \frac{d^3\p\ }{(2\pi)^3}  [ \tilde{B}_L + \tilde{C}_L ].
\end{eqnarray}

\noindent Substituting into $I_1$ the expressions (\ref{apLandau})
for $ \tilde{B}_L, \tilde{C}_L, \tilde{D}_L $ we obtain

\begin{equation}
I_1 = \int \frac{|\p |^2 d|\p |}{4 \pi^2} \frac{dn}{dE} \left[
\frac{|\p  |^2 |\q |^2}{3m^2} + \frac{|\Delta|^4}{\epsilon^2 E^2} \left( 1 +
\frac{a}{2} I_s \right) q_0^2 \right],
\end{equation}

\noindent where \[\ I_s =
\int_{-1}^{1} \frac{ds}{s-a -i \delta \ \mathrm{sgn}(\epsilon ) },
\hspace{0.5cm} a=\frac{q_0 E m}{\epsilon |\p | |\q |} .\] 
The integral $ I_s $ has a pole at $ a $, if $ -1<a<1 $. The sign of
$\epsilon $ appears in the denominator of $I_s$, because the
infinitesimal imaginary part $ i \delta $ is multiplied by $a$ whose
sign depends only on the sign of $\epsilon $, the rest of the
parameters in $a$ being always positive. Performing the $s$-integration
yields

\begin{equation}
I_s = \ln{ | \frac{1-a}{1+a} | } + i \pi \ \mathrm{sgn}(\epsilon ) \ \theta (1-|a|) .    \label{Is}
\end{equation}

\noindent Similar expressions can be found for $I_2$ and $I_3$.
 Applying now approximation (\ref{fermi}) we find 

\begin{eqnarray}
I_1 &=& (\frac{3 \rho_0 }{4 m}) \left( \frac{1}{3} J |\q |^2 -(H_r^{(1)}(c)
+ i H_i^{(1)}(c)) q_{0}^{2}/v_{F}^{2})
 \right)  \nonumber  \\
I_2 &=& i ( \frac{3 \rho_{0}}{2 m v_{F}^{2}} ) \:
q_{0} \: (H_r^{(2)}(c) + i H_i^{(2)}(c) ) \nonumber \\
I_3 &=& ( \frac{3 \rho_{0}}{2 m v_{F}^{2}}) \ (H_r^{(3)}(c) + i H_i^{(3)}(c) ) \nonumber \\
\end{eqnarray}

\noindent where 

\begin{eqnarray}
J &=& | \Delta | \int dy \frac{dn}{dE} \nonumber \\
H_r^{(n)}(c) &=&  -| \Delta |  \int dy \frac{dn}{dE} \frac{y^{2(n-2)}}{
(y^2+1) } \left[ 1 - \frac{c}{2} \frac{ \sqrt{y^2+1} }{y} \ln{ | \frac{y + c
\sqrt{y^2+1}}{y-c \sqrt{y^2+1}} |} \right] \nonumber \\
H_i^{(n)}(c) &=& -\frac{|\Delta| \pi c}{2} \int dy \frac{dn}{dE}
\frac{y^{2(n-2)}}{|y| \sqrt{1+y^2}} \theta ( \frac{|y|}{ \sqrt{1+y^2}} -
|c|) \nonumber \\
c &=& \frac{1}{v_F} \frac{q_0}{|\q |}. \label{defLandau}
\end{eqnarray}

\noindent We observe that, for $|c| \geq 1$, $H_i^{(n)}=0$
, so that the cut  in the $H^{(n)}$ functions is over the region 
$| c | \leq 1 $. It is noteworthy that the function $H^{(1)}$, which 
multiplies the term in $q_{0}^{2}$, is, apart from a factor of 4 and 
slight changes in notation, exactly the same as the corresponding 
function $H$ of \cite{Dominic}.

Substituting the expressions which we have found for $I_1, I_2, I_3$
back into the second trace, making approximation (\ref{M}),
 and Fourier transforming the $ \tilde{
\theta} $ functions wherever this is possible, we can bring the effective action to the following form

\begin{eqnarray}
 S_{\mathit{eff}} & = & \int d^4x \left( -\frac{3 \rho_{0}}{8m} \right) \left[   \frac{4m}{3} \left(
\dot{\theta} + [1+J ]\frac{ ( \nabla \theta )^2}{4m} \right) -
 \frac{R }{2 v_{F}^{2}}  \left( \dot{ \theta } + \frac{(\nabla
\theta)^2}{4m} \right)^2  \right. \nonumber \\
 && - \int \frac{d^4q}{(2 \pi)^4} \tilde{ \theta } (q) \tilde{ \theta }
(-q) \frac{ q_0^2}{v_{F}^{2} } (H_r^{(1)}(c) +i H_i^{(1)}(c)) \nonumber \\ 
&& - i \int \frac{d^4q}{(2 \pi )^4} \frac{d^4k}{(2
\pi)^4} \tilde{ \theta }(q) \tilde{ \theta}(k)  \tilde{ \theta }(-q-k)
  2q_0 \frac{
(\k .\q + \k ^2 )}{4m} \frac{1}{v_{F}^{2}} (H_r^{(2)}(c) + i H_i^{(2)}(c)) 
\nonumber \\
&&  + \int \frac{d^4q}{(2 \pi)^4} \frac{d^4k}{(2 \pi)^4} \frac{d^4r}{(2\pi)^4} \tilde{ \theta }(k) \tilde{ \theta }(r)
  \tilde{ \theta }(q-k) \tilde{ \theta }(-q-r )  \nonumber \\
&&  \left. \frac{(-\k ^2 + \k .\q )}{4m} \frac{(\r ^2 + \r .\q )}
{4m}\frac{1}{v_{F}^{2}} ( H_r^{(3)}(c) + i H_i^{(3)}(c)) \right] . \label{Leff}
\end{eqnarray}

Consider now the effect of the Landau terms $J$ and $H^{(n)}$ in (\ref{Leff}).
The simplest of these is $J$, which is a (temperature-dependent) constant.
Thus its contribution is still local but its presence does break Galilean 
invariance, since it disrupts the invariant combination ``$\dot{\theta} + 
(\nabla \theta )^2 /4m $'' in the first term in square brackets in (\ref{Leff}).
Actually, as far as this single term is concerned, Galilean invariance can be 
formally restored by modifying the first $\dot{\theta}$ term so as to read
$[1+J]\dot{\theta}$~; this does not affect the equations of motion.

The effect of the $H^{(n)}$ functions is - depending on their numerical 
magnitudes - potentially much more drastic. Most importantly, these functions 
cannot be expanded about $q=0$, since they depend explicitly on the ratio 
$c = q_{0}/v_{F} | {\bf q} |$. Thus the terms involving the $H^{(n)}$'s in 
(\ref{Leff}) cannot be represented as any Lagrangian local in coordinate 
 space. {\em A fortiori} then, these terms also break Galilean
invariance. They also contain imaginary 
parts, of course, corresponding to the physical `` Cerenkov '' radiation 
process in the heat bath \cite{Tsuneto,Dominic}. In the next section we discuss how 
the equation of motion for $\theta$, which follows from (\ref{Leff}), may be 
approximated by a more manageable form, which may be given a physical 
interpretation. 

\section{ Approximate equation of motion for $\theta $ and its
interpretation }

\subsection{The $\theta $-propagator}

Let us begin by considering the terms in (\ref{Leff}) which are
quadratic in $\tilde{\theta}$, from which we can read off the
momentum-space propagator for $\theta$. The first point to note is
that, since $H_i^{(1)}$ is an odd function of $q_0$, it contributes
nothing to the action $S_{\mathit{eff}}$. To see the effect of damping we need
to consider the equation of motion for $\tilde{\theta}$. This has the
form 

\begin{equation}
\tilde{ \theta } = S_{G} (q_0, |\q | ) \left[ \int \ \tilde{\theta} ^2
 \ H_{r,i}^{(2)} (c), \int \int \ \tilde{ \theta } ^3  \ H_{r,i}^{(3)} (c) \right]
\label{equation of motion}
\end{equation}

\noindent where we have written the non-linear terms on the right hand
side in symbolic form and where the $\theta $-propagator is given by

\begin{equation}
S_{G} ( q_0, |\q | ) = \left\{ \q ^2 \left[ \frac{c^2 R }{2} -
\frac{1+J}{3} + c^2 \left( H_r^{(1)} (c) + i H_i^{(1)} (c) \right)
\right] \right\}^{-1} 
\label{theta propagator}
\end{equation}

\noindent At $T=0$ we have $R=2$ and $J=H_r^{(1)}=H_i^{(1)}=0$, so
that $S_{G}$ has poles at $c= \pm \frac{1}{\sqrt{3}} $, as
expected for the Bogoliubov-Anderson mode \cite{Anderson,Bogoliubov}.

Consider now what happens as $T$ moves away from zero. $\frac{R}{2}$
and $J$ deviate slowly from $1$ and $0$ respectively (see
fig.\ref{fig.1}a, fig.\ref{fig.1}b ) while the quantities $c^2 H_r^{(1)}$ and $c^2 H_i^{(1)}$ are very small compared to
the main term $- \frac{1}{3} $ (see fig.\ref{fig.2a}a, fig.\ref{fig.2b}b ). This suggests
that it should be a good approximation to replace the complicated
$c$-dependence of the denominator in (\ref{theta propagator}) by a
quadratic function, which amounts to replacing the cut structure of
(\ref{theta propagator}) by two poles - the same two that were present
at $T=0$ but now at slightly different positions. In particular, the
poles will now be shifted off the real axis, due to the presence of
the $H_i^{(1)}$ function. It is important to note that $H_r^{(1)}$ is
an even function of real $c$ while $H_i^{(1)}$ is an odd function of
real $c$. It is then clear that the poles of (\ref{theta propagator})
are approximately at $ c = \pm c_r - i \gamma $, where
$ c_r = \pm \frac{1}{\sqrt{3}} $ and $ \gamma \ll c_r $; both of the
approximate poles are in the lower half plane, as expected, since we
are considering a retarded propagator, which is required for the study of damping.  
 
To find expressions for $c_r$ and $\gamma $ we write $ H_i^{(1)} (c) = c \
\tilde{H}_i^{(1)} (c^2)$, so that, after the approximation which
follows, we do not lose the property that $ H_i^{(1)} $ is an odd
function of $c$. Now, since the functions $H_r^{(1)}$ and
$H_i^{(1)}$ are small compared to the regular term, we can make an approximation similar to
the one which leads to the well-known Breit-Wigner formula;
we expand the quantities $c^2 H_r^{(1)} (c^2)$ and $c^2 \tilde{H}_i^{(1)} (c^2)$
which appear in (\ref{theta propagator}) about $ c^2=\frac{1}{3} $
keeping only the lowest order term. Thus we arrive at 

\begin{equation}
S_{G} (q_0, |\q |) \approx  \left\{ \q ^2 \frac{R}{2} \left[ c^2 -
\frac{2(1+J)}{3R} + \frac{ 2 \bar{H}_r^{(1)}}{3 R} + i \frac{2c \
\bar{H}_i^{(1)}}{\sqrt{3} R} \right] \right\}^{-1} \label{Breit 1}
\end{equation}

\noindent where the bar denotes quantities evaluated at
$c=\frac{1}{\sqrt{3}}$. (\ref{Breit 1}) exhibits the desired quadratic behaviour and has poles at $c=\pm c_r
- i \gamma $, where \[ c_r = \sqrt{ \frac{2(1+J)}{3R} - \frac{ 2 \bar{H}_r^{(1)}}{3 R}}
\hspace{1cm} \gamma = \frac{ \bar{H}_i^{(1)} }{\sqrt{3} R} \]
and $\gamma ^2 $ has been neglected. $c_r$ and $\gamma $, as functions
of temperature, are plotted in fig.\ref{fig3}a and fig.\ref{fig3}b
respectively. They both have the expected behaviour, which reassures
us about the validity of our approximation. In terms of $c_r$ and $\gamma $
we may recast (\ref{Breit 1}) as follows 
\begin{equation}
S_{G} (q_0, |\q |) \approx \left\{ \frac{R}{2 v_F^2} \left[ q_0^2 - c_r^2 \
v_F^{2} \ \q ^2 + 2 i \ \gamma \ v_F \ q_0 \ |\q | \right] \right\}^{-1}
\label{Breit 2}
\end{equation}
\noindent or equivalently as

\begin{equation}
S_{G} \approx \frac{v_F}{ c_r \ R \ |\q |} \ \left\{ \frac{1}{q_0 - c_r \ v_F \
|\q | + i \gamma \ v_F \ |\q |} - \frac{1}{q_0 + c_r \ v_F \
|\q | + i \gamma \ v_F \ |\q |} \right\} 
\label{Breit 3}
\end{equation}

\noindent which manifestly has the correct form for a bosonic
propagator with damping, the line-width having an explicit $|\q
|$-dependence. 

At this stage we have to point out that the alternative approximate
form for $S_{G}$ given in \cite{Dominic} is incorrect, since it fails
to exhibit the important property (traceable to $H_i^{(1)}$ ) that the
imaginary part should be an odd function of $q_0$. The pole of $S_{G}$
at $c = - c_r - i \gamma $ was overlooked in \cite{Dominic} and
instead a pole at $c= - c_r + i \gamma $ was used, which is actually
on a Riemann sheet distant from the physical boundary. This mistake
had the effect of replacing the $2 i \ \gamma \ v_F \ q_0 \ |\q | $
term in (\ref{Breit 2}) with something proportional to $ \q ^2 $. Such
a form then does, of course, have a simple local Fourier transform. In
contrast (\ref{Breit 2}) does not but, as noted in the start of the
section, the odd-$q_0$ piece will integrate to zero in
$S_{\mathit{eff}}$. Damping can only be manifested in the equation of motion
and in this case, when included, it will lead to a non-local term in the effective wave-operator (the
transform of $S_{G}^{-1}$). Nevertheless, we do have a simple and
reliable expression (\ref{Breit 2}) for $S_{G}$ in momentum space, 
 which shows a particular form of damping and a predicted
temperature-dependence for the line-shape. As noted in the
Introduction, the non-local form of $S_{G}^{-1}$ appears to present a
fundamental difficulty for the inclusion of electromagnetism.  

We now turn to the treatment of the other terms on the right hand side
of (\ref{equation of motion}).  

\subsection{The remaining Landau terms involving $H^{(2)}$ and $H^{(3)}$.}

The fact that $S_{G}$ is so sharply peaked around $q_0 \approx \pm
\frac{1}{ \sqrt{3}} v_F |\q | $ suggests that, as far as the dynamics
of $\theta $ described by (\ref{equation of motion}) are concerned, it
should be a good approximation to evaluate the $c$-even functions $H_r^{(2)}$ and
$H_r^{(3)}$ at $c^2=\frac{1}{3}$. The corresponding imaginary parts
are very small, $c$-odd functions, as $ H_{i}^{(1)} $ was, and we shall from now on simply ignore them.
 In this case all the terms in (\ref{Leff}) become local and the Fourier
transforms to $x$-space can all be done, leading to the effective Lagrangian
   
\begin{eqnarray}
{\cal{L}}_{R+\bar{L}_i} &=& -M'(T) [\dot{\theta} +
\frac{ (\nabla \theta)^2}{4m}] + \frac{N(T)}{4} [\dot{\theta} + \frac{(\nabla \theta)^2}{4m}]^2 \nonumber\\
&+& \frac{3 \rho_0}{8mv^2_F} \left(\dot{\theta}^2 \bar{H}^{(1)}_r + 2
\dot{\theta} \frac{(\nabla \theta)^2}{4m} \bar{H}^{(2)}_r +
\frac{[(\nabla \theta)^2]^2}{(4m)^2} \bar{H}_r^{(3)} \right ), \label{full}
\end{eqnarray}

\noindent where \[ M'(T) = \frac{\rho _0}{2} \ (1+J) \] is (\ref{M})
as modified by the contribution from the Landau term, $ \bar{H}_r^{(j)} \equiv H_r^{(j)}
(c^2=\frac{1}{3}) \ (j=1,2,3) $ and we have modified the $\dot{\theta}$ term as
suggested after (\ref{Leff}). 
We note that the first part of ({\ref{full}), not involving the
$\bar{H}_r^{(j)} \ (j=1,2,3) $ can be written as a TDNLSL just as in
section 5, with a potential \[ V=\frac{(\rho -2 M')^2}{2N} . \]
Choosing a convenient normalisation, the equation of motion for $\theta$ is then
\begin{eqnarray}
&~& \frac{\partial}{\partial t} \left [ \rho - \frac{3 \rho_0}{2mv^2_F} (\dot{\theta} \bar{H}_r^{(1)} + \frac{(\nabla \theta)^2}{4m} \bar{H}_r^{(2)} ) \right ] \nonumber\\
&+& {\bf \nabla} . \left \{ \frac{\nabla \theta}{2m} \left [\rho - \frac{3 \rho_0}{2mv^2_F} \left ( \dot{\theta} \bar{H}_r^{(2)} + \frac{(\nabla \theta)^2}{4m} \bar{H}^{(3)}_r \right ) \right ] \right \} = 0 \label{cty}
\end{eqnarray}
\noindent where
\begin{equation}
\rho = 2M' - N [ \dot{\theta} + \frac{(\nabla \theta)^2}{4m}].
\end{equation}
It is tempting to interpret the quantity
\begin{equation}
\rho_n = - \frac{3 \rho_0}{2mv^2_F} \left ( \dot{\theta} \bar{H}_r^{(1)} + \frac{(\nabla \theta)^2}{4m} \bar{H}^{(2)}_r \right ) \label{rho}
\end{equation}
in (\ref{cty}) as an additional ``density'', not described by the NLSL.  On this interpretation, we would expect the corresponding additional term inside the divergence in (\ref{cty}) to represent the associated ``current'', which would require
\begin{equation}
\rho_n {\bf v}_n = - \frac{3 \rho_0}{2mv^2_F} \left ( \dot{\theta}
\bar{H}_r^{(2)} + \frac{(\nabla \theta)^2}{4m} \bar{H}_r^{(3)} \right
) \frac{\nabla \theta}{2 m}. \label{rhov}
\end{equation}
Equations (\ref{rho}) and (\ref{rhov}) are compatible, if
\begin{equation}
{\bf v}_n = \frac{\nabla \theta}{2m^*} \label{vely}
\end{equation}
\begin{equation}
m^* = (\bar{H}^{(1)}_r / \bar{H}^{(2)}_r )m  \label{m*}
\end{equation}
\begin{equation}
\bar{H}^{(3)}_r \bar{H}^{(1)}_r = (\bar{H}^{(2)}_r)^2. \label{compat}
\end{equation}
In this case, (\ref{cty}) becomes the continuity equation of the two-fluid model
\begin{equation}
\frac{\partial}{\partial t} (\rho + \rho_n) + {\bf \nabla} .(\rho {\bf v} + \rho_n {\bf v}_n) =0
\end{equation}
\noindent where $\rho$ is the superfluid density and  
the superfluid velocity is ${\bf v}= \nabla \theta /2m$ as usual.

The crucial requirement is (\ref{compat}). Fig.\ref{fig4}a shows the ratio $\bar{H}_r^{(3)} \bar{H}_r^{(1)}/(\bar{H}_r^{(2)})^2$ over the range of $b$ we are considering, and it can be seen that (\ref{compat}) is quite well satisfied. Fig.\ref{fig4}b 
shows the ratio $\bar{H}_r^{(1)}/\bar{H}_r^{(2)}$ as a function of
$b$, from which we see that $m^*$ in (\ref{m*}) varies quite rapidly
with $T$ at low $T$, and then approaches the constant value $5m$.

It is easy to show that there is a TDNLSL for the normal fluid from
which the continuity equation for the normal fluid can be derived as
an equation of motion. One can then use the ``minimal coupling''
prescription to take electromagnetism into account. 

In summary, we have argued that the effect of the Landau terms
$H^{(n)}$ can be approximately incorporated by treating them as a
constant, evaluating them at $c^2=1/3$. Considering the resulting
effective Lagrangian, it seems to us to describe a two-fluid model in
which the superfluid component could be described by a TDNLSL with a
temperature dependent potential, and the normal component by another
TDNLSL with an appropriately chosen potential and a
 definite velocity predicted by (\ref{vely}). 

\vspace*{1cm}

\begin{center}
{\large {\bf Acknowledgements }}
\end{center}

 We wish to thank Cesar Fosco, Marcelo Hott and Alexei Tsvelik for valuable discussions.
 I.J.R.A. gratefully acknowledges support from the Antorchas
Foundation (Argentina) and the British Council, which enabled him to
visit the Instituto Balseiro, Bariloche, where part of this work was
completed. D.J.L. and G.M. gratefully acknowledge financial support
from P.P.A.R.C. (UK).

\vspace*{1cm}

\newpage


\newpage

\begin{figure}
\begin{center}
\includegraphics[totalheight=0.85\textheight=10]{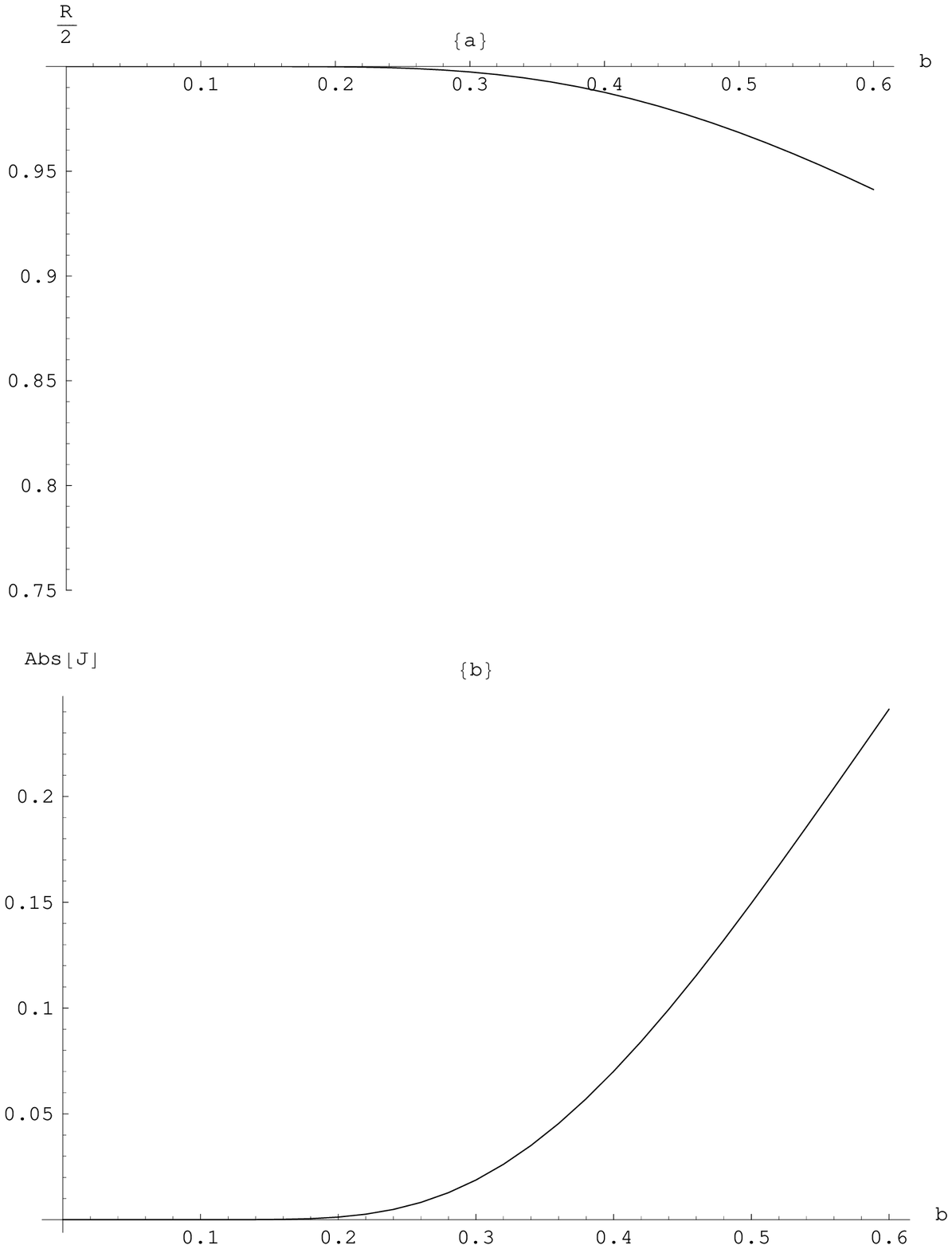}
\caption{(a): $\frac{R(T)}{2}$ is very close to $1$ for any
temperature in the region which is considered here, $b =
\frac{T}{T_c} \in [0,0.6]$. (b): $|J(T)|$ is much smaller than $1$ for
any temperature in the same region. \label{fig.1}}
\end{center}
\end{figure}

\begin{figure}
\begin{center}
\includegraphics[totalheight=0.85\textheight=10]{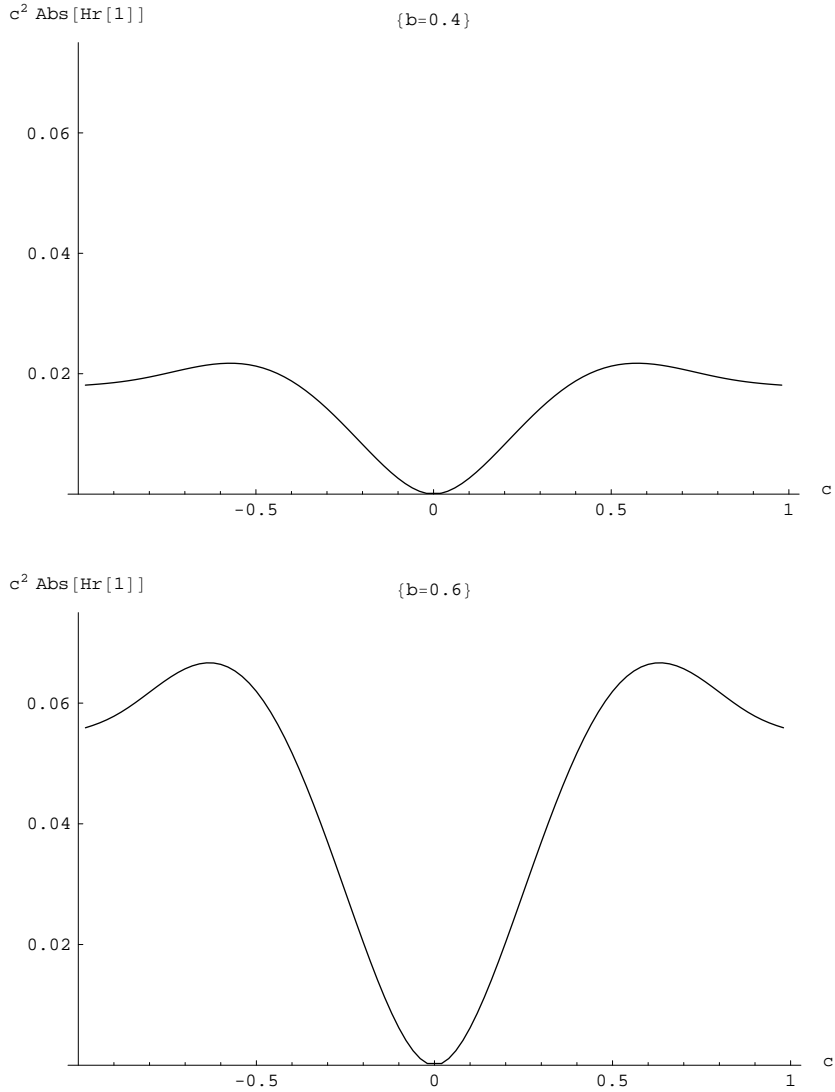}
\caption{(a): $c^2 |H_r^{(1)}|$ is negligible compared to  $\frac{1}{3} \approx
0.333$, for any temperature $b \in [0,0.6] $ and $|c| \leq 1$. For
$|c| \geq 1 $, it can be shown that
 the $c$-even function $c^2 | H_r^{(1)} | $ is always much smaller
than $\frac{1}{3}$.
 \label{fig.2a} }
\end{center}
\end{figure}

\setcounter{figure}{1}
\begin{figure}
\begin{center}
\includegraphics[totalheight=0.85\textheight=10]{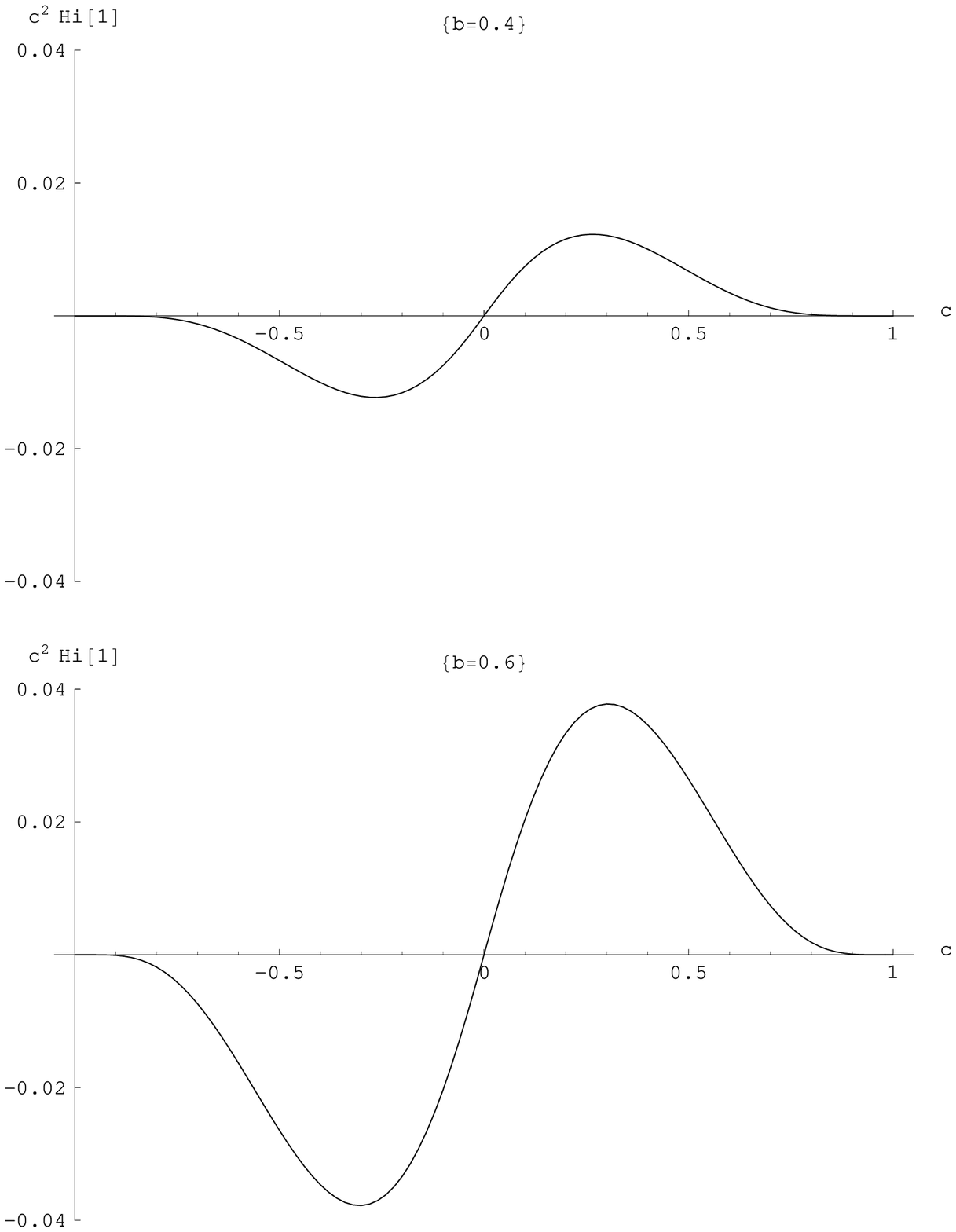}
\caption{(b): $c^2 |H_i^{(1)}|$ is negligible compared to $\frac{1}{3}
\approx 0.333$, for any temperature $b \in [0,0.6] $ and $|c| \leq
1$. For $|c| \geq 1$, $H_i^{(1)}=0$. \label{fig.2b} }
\end{center}
\end{figure}

\begin{figure}
\begin{center}
\includegraphics[totalheight=0.90\textheight=12]{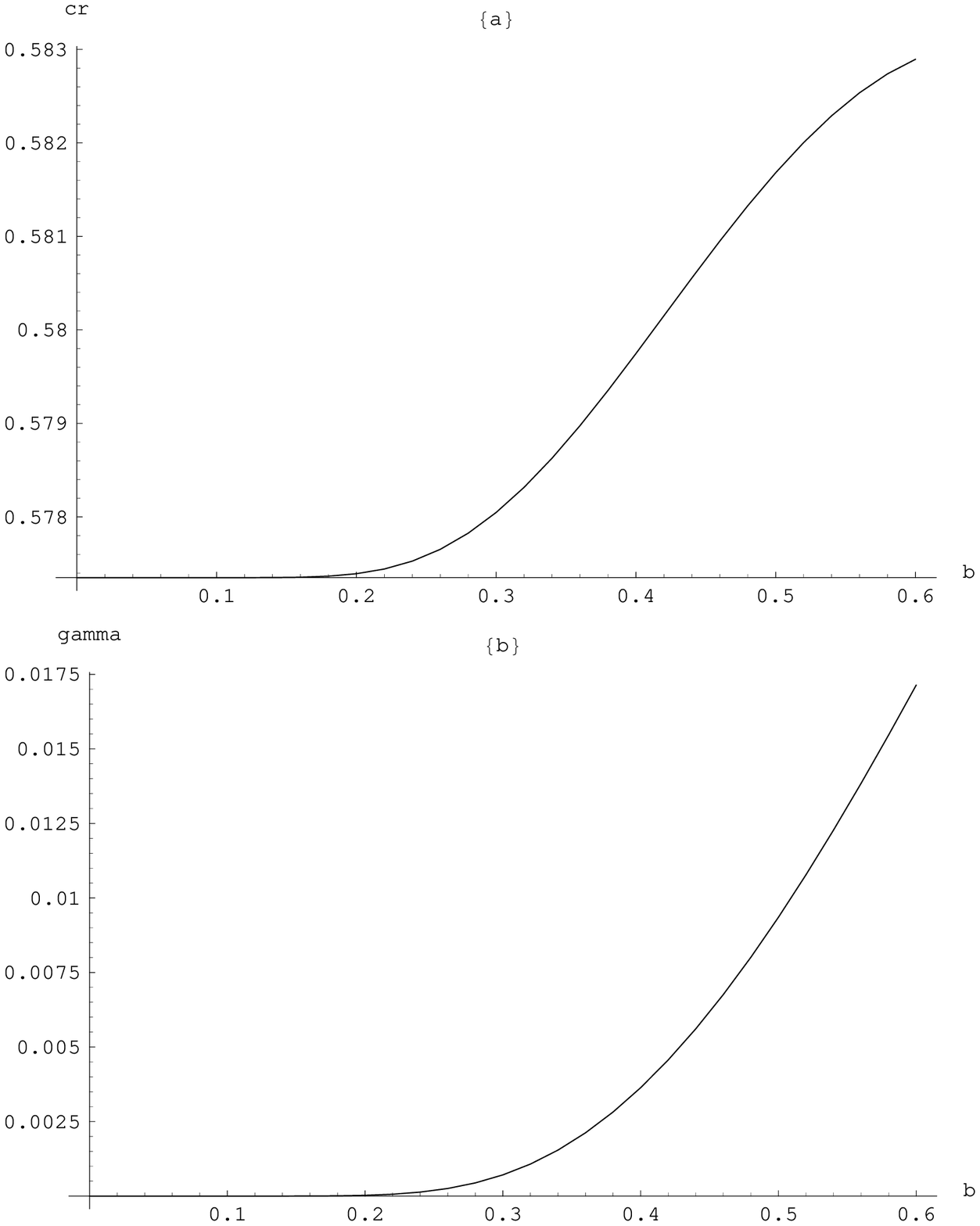}
\caption{(a): $c_r$ is very close to the original location of the
absolute value of the poles $\frac{1}{\sqrt{3}} \approx 0.577 $, for
any temperature in the considered region. (b): The damping $\gamma $ is negligible
compared to the real part of the pole, for any temperature in the considered
 region. \label{fig3} }
\end{center}
\end{figure}

\begin{figure}
\begin{center}
\includegraphics[totalheight=0.90\textheight=12]{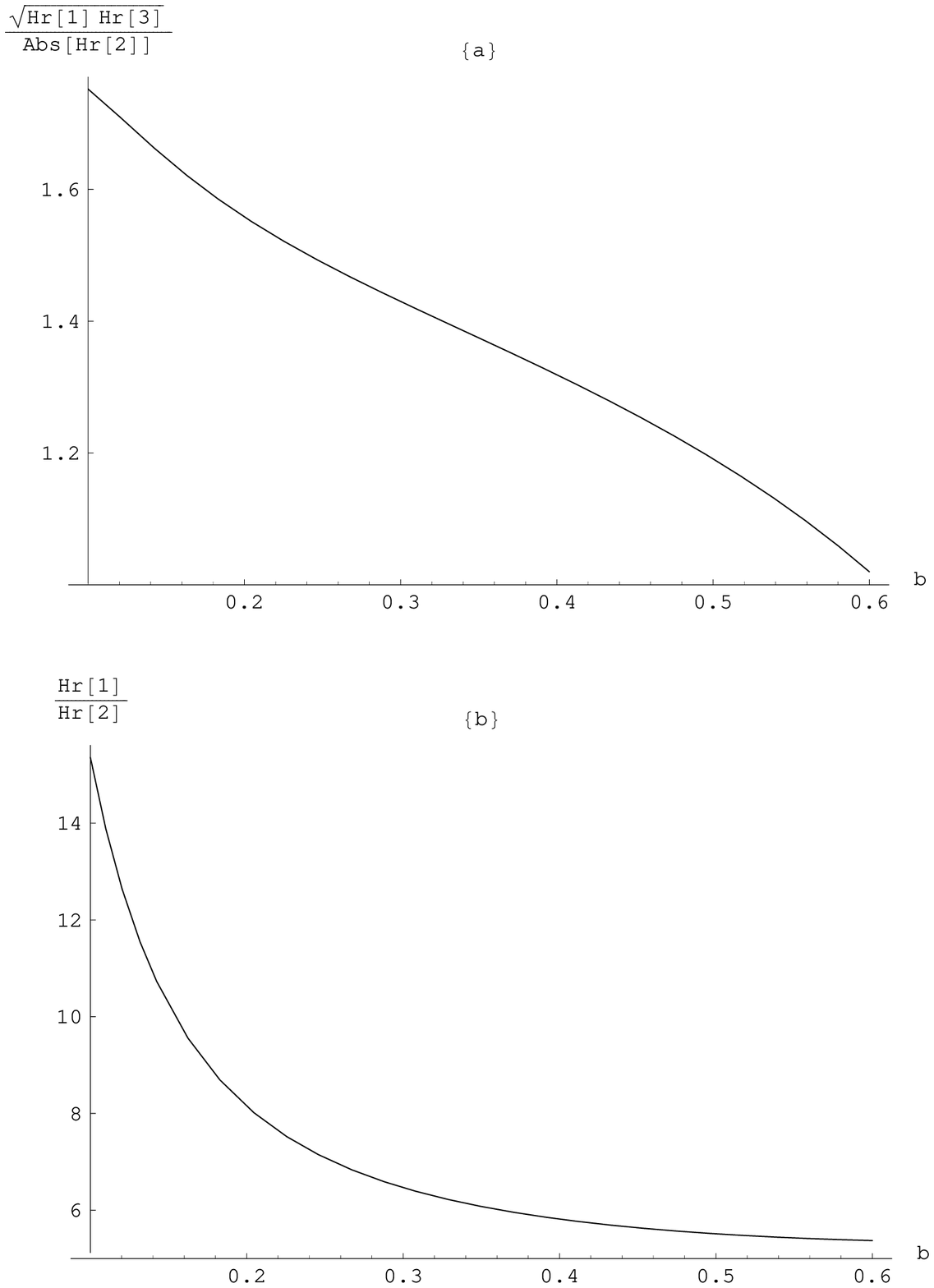}
\caption{(a):  As the temperature rises $
\sqrt{\bar{H}_r^{(1)} \bar{H}_r^{(3)} }/ | \bar{H}_r^{(2)} |$
approaches 1. (b): As the temperature rises $ \bar{H}^{(1)}_r / \bar{H}^{(2)}_r = m^* / m $ approaches 5. \label{fig4} }
\end{center}
\end{figure}

\end{document}